\documentclass[
srcltx,aps,floats,amssymb,latexsym,eucal,url,prl,hyperref]{revtex4}

\DeclareFontFamily{OT1}{rsfs}{}
\DeclareFontShape{OT1}{rsfs}{m}{n}{ <-7> rsfs5 <7-10> rsfs7 <10-> rsfs10}{}
\DeclareMathAlphabet{\mycal}{OT1}{rsfs}{m}{n}

\renewcommand{\cal}{\mycal}
\newcommand{\Isot}{\text{Iso}^\uparrow(\hyp,b)}
\newcommand{\Iso}{\text{Iso}(\hyp,b)}

\newcommand{\A}{A}  
  \newcommand{\ce}{\beta}

\newcommand{\mnoteprzelicz}[1]{}
\renewcommand{\lrcorner}{\rfloor} 

\newcounter{mnotecount}[section]

\newcommand{\FS}       
                  {F}

\newcommand{\HS} 
       {H_{\mbox{\scriptsize volume}}}

\newcommand{\ren}{\mbox{\scriptsize normalised}}
\newcommand{\HSren} 
       {{\HS^{\ren}}}

\newcommand{\ptc}[1]{}

\newcommand{\cU}{{\cal U}}

\newcommand{\eq}[1]{(\ref{#1})}

      \newcommand{\be}{\begin{equation}}
      \newcommand{\ee}{\end{equation}}
\newcommand{\eea}{\end{eqnarray}}
\newcommand{\bea}{\begin{eqnarray}}

      \newfont{\msa}{msam10 scaled\magstep1}

\def\Reals{{\Bbb R}}

\newcommand{\rd}{\,{ d}} 

\newcommand{\R}{\Reals} 

\newcommand{\hyp}{{\cal S}}

\newcommand{\bmetric}{{b}} 

\newcommand{\Kbasis} {{K}} 

\newcommand{\ourU}{{\mathbb U}}

\newcommand{\hhyp}{\,\,\widehat{\!\!\hyp}\,} 

\newcommand{\znabla} {\mathring{\nabla}} 

\newcommand{\mn} {M} 
\newcommand{\cM}{{\mycal M}} 
\newcommand{\cKS}{{\mycal K}_{\hyp^{\perp}}} 
\newcommand{\Eq}[1]{Equation~\eq{#1}}

\newcommand{\hb}{{\hat b}}

\begin{document}

\draft


\title{The Hamiltonian mass of asymptotically anti-de Sitter space-times}

\author{Piotr T.\ Chru\'sciel\footnote{ Supported in part by the Polish
    Research Council grant KBN 2 P03B 073 15. E--mail:
    \texttt{chrusciel@univ-tours.fr}} and Gabriel Nagy\footnote{
    Supported by a grant from R\'egion Centre. E--mail:
    \texttt{nagy@gargan.math.univ-tours.fr}}} \affiliation{ D\'epartement
    de Math\'ematiques, Facult\'e des Sciences, Parc de Grandmont,
    F-37200 Tours, France}

\begin{abstract}
  We give a Hamiltonian definition of mass for
  spacelike hypersurfaces in space-times with metrics which are
  asymptotic to the anti-de Sitter one, or to a class of
  generalizations thereof. We present the results of~\cite{ChNagy2}
  which show that our definition provides a geometric invariant for a
  spacelike hypersurface $\hyp$ embedded in a space-time
  $({\cM},g)$. Some further global invariants are also given.
\end{abstract}

\pacs{11.30-j, 04.20.-q, 02.40.Ma}

\maketitle


Let $\hyp$ be an $n$-dimensional spacelike hypersurface in a
$n+1$-dimensional Lorentzian space-time $(\cM,g)$. Suppose that $\cM$
contains an open set $\cU$ which is covered by a finite number
of
coordinate charts $(t,r,v^A)$, with $r\in[R,\infty)$, and with $(v^A)$
--- local coordinates on some compact $n-1$ dimensional manifold
$\mn$, such that $\hyp\cap \cU=\{t=0\}$. Assume that the metric $g$
approaches a background metric $b$ of the form
\begin{eqnarray}\nonumber
  b= -{ a^{-2}( r)} dt^2 +  a^2( r) dr^2
  +r^2 h\;, \\
   h=h_{AB}( v^C)d v^A d v  ^B\;,
  \label{eq:a1}
\end{eqnarray}
with $a(r)=1/\sqrt{r^2/\ell^2+k}$, where $h$ is a Riemannian metric on
$\mn$, $k$ is a constant, and $\ell$ is a strictly positive constant
related to the cosmological constant $\Lambda$ by the formula
$2\Lambda = -{n(n-1)/
\ell^2}$ (somewhat more general metrics are considered in~\cite{ChNagy2}). For
example, if $h$ is the standard round metric on $S^2$ and $k=1$, then
$b$ is the anti-de Sitter metric. To make the approach rates precise
it is convenient to introduce an orthonormal frame for $b$,
\begin{eqnarray}
  \label{frame}
e_0 = a (r)\partial_t\;, \qquad e_1 = {1 \over a(r) } \partial _r\;,
\quad
  e_{\A} = \frac 1r \ce_{\A}\;,
\end{eqnarray}
with $\ce_A$ --- an $h$-orthonormal frame on $(M,h)$, so that
$b_{ab}=b(e_a,e_b)=\eta_{ab}$ --- the usual Minkowski matrix
diag$(-1,+1,\cdots,+1)$. We then require that the frame components
$g_{ab}$ of $g$ with respect to the frame \eq{frame}
satisfy\footnote{The summation convention is used throughout. We use
    Greek indices for coordinate components and lower-case Latin
    indices for the tetrad ones; upper-case Latin indices run from $2$
    to $n$ and are associated either to coordinates or to frames on
    $\mn$, in a way which should be obvious from the context. Finally
  Greek \emph{bracketed} indices $(\alpha)$ \emph{etc.} refer to
  objects defined outside of space-time, such as the Killing algebra
  $\cKS$, or an exterior embedding space, and are also summed
over.}
\begin{equation}
  \label{C4}
  e^{ab}= O(r^{-\beta})\;, \quad e_a(e^{bc}) = O(r^{-\beta})\;, \quad
  b_{ab}e^{ab}= O(r^{-\gamma})\;,
\end{equation}
where $e^{ab}=g^{ab}-b^{ab}$, with
\begin{equation}
  \label{C5}
  \beta>n/2\;, \qquad \gamma > n\;.
\end{equation}
(The $n+1$ dimensional Schwarzschild-anti de Sitter metrics
\eq{Kottler2} satisfy \eq{C4} with $\beta=n$, and with
$\gamma=2n$. The boundary conditions \eq{C4}-\eq{C5}, and the
overall approach here stem from~\cite[Section~5]{ChruscielSimon}.)
We stress that we do {\em not} assume
existence of global frames on the asymptotic region: when $M$ is not
parallelizable, then  conditions \eq{C4}--\eq{C5} should be understood as
the requirement of {\em
existence of a covering of $M$ by a finite number of  open sets
${\mycal O}_i$ together with frames defined on
$[R_0,\infty)\times {\mycal O}_i$ satisfying \eq{C4}-\eq{C5}.}
Suppose further, for simplicity, that $g$ satisfies the vacuum
Einstein equations with a cosmological constant,
\begin{equation}
  \label{Ee} R_{\mu\nu}-\frac{g^{\alpha \beta} R_{\alpha \beta}}{2}
  g_{\mu\nu} = -\Lambda g_{\mu\nu} \;,
\end{equation}
similarly for $b$. (The existence of a large family of such $g$'s
follows from the work in~\cite{Friedrich:adS,Kannar:adS}.)  A
geometric Hamiltonian analysis~\cite{ChAIHP,ChNagy2}, based on the
formalism of~\cite{KijowskiTulczyjew}, leads to the following
expression for the Hamiltonian associated to the flow of a vector
field $X$, assumed to be a Killing vector field of the background
$b$:\footnote{The integral over $\partial \hyp$ should be understood
by a limiting process, as the limit as $R$ tends to infinity of
integrals over the sets $t=0$, $r=R$. $d S_{\alpha\beta}$ is defined
as $\frac{\partial}{\partial x^\alpha}\lrcorner
\frac{\partial}{\partial x^\beta}\lrcorner \rd x^0 \wedge\cdots
\wedge\rd x^{n} $, with $\lrcorner$ denoting contraction; $g$ stands
for the space-time metric unless explicitly indicated otherwise.
Square brackets denote antisymmetrization with an appropriate
numerical factor ($1/2$ for two indices), and $\znabla$ denotes
covariant differentiation \emph{with respect to the background metric
$b$}.}
\begin{eqnarray}
  m(\hyp,g,b, X)&= &\frac 12 \int_{\partial\hyp}
 \ourU^{\alpha\beta}dS_{\alpha\beta}\;,
\label{toto}
\end{eqnarray}
\begin{eqnarray}
  \ourU^{\nu\lambda}&= &
{\ourU^{\nu\lambda}}_{\beta}X^\beta
+ \frac 1{8\pi} \left(\sqrt{|\det
g_{\rho\sigma}|}~g^{\alpha[\nu}\right.
\left. -\sqrt{|\det b_{\rho\sigma}|}~
b^{\alpha[\nu}\right)\delta^{\lambda]}_\beta \znabla_{\alpha}{X^\beta}
\ ,\label{Fsup2new}
\\ {\ourU^{\nu\lambda}}_\beta &= & \displaystyle{\frac{2|\det
  \bmetric_{\mu\nu}|}{ 16\pi\sqrt{|\det g_{\rho\sigma}|}}}
g_{\beta\gamma}\znabla_{\kappa}\left(e^2 g^{\gamma[\nu}g^{\lambda]\kappa}\right)
\;,\label{Freud2.0} 
\\
  \label{mas2}
e&=& {\sqrt{|\det
g_{\rho\sigma}|}}/{\sqrt{|\det\bmetric_{\mu\nu}|}}\; .
\end{eqnarray}
It is appropriate to make a few comments here: {In~\cite{ChAIHP}
the starting point
of the analysis is the Hilbert Lagrangian for vacuum Einstein gravity,
${\mathcal{L}}= \sqrt{- \det g_{\mu\nu}}{g^{\alpha \beta} R_{\alpha
\beta}}/{16\pi}. $ With our signature $(-+\cdots+)$ one needs to
repeat the analysis in~\cite{ChAIHP} with $\mathcal{L}$ replaced by
${\sqrt{- \det g_{\mu\nu}}}\left(g^{\alpha \beta} R_{\alpha \beta}
-2\Lambda\right)/{16\pi}$, and without making the assumption $n+1=4$
done there. The final expression for the Hamiltonian \eq{toto} ends up
to coincide with that obtained when $\Lambda=0$ and
$n+1=4$. Equation~\eq{Fsup2new} coincides with that derived
in~\cite{ChAIHP} except for the term $-\sqrt{|\det
b_{\rho\sigma}|}~b^{\alpha[\nu}\delta^{\lambda]}_\beta
\znabla_{\alpha}{X^\beta}$. This term does not depend on the metric
$g$, and such terms can be freely added to the Hamiltonian because
they do not affect the variational formula that defines a
Hamiltonian. {}From an energy point of view such an addition
corresponds to a choice of the zero point of the energy. We note that
in our context $m(\hyp,g,b,X )$ would not converge if the term
$-\sqrt{|\det b_{\rho\sigma}|}~b^{\alpha[\nu}\delta^{\lambda]}_\beta
\znabla_{\alpha}{X^\beta}$ were not present in \eq{Fsup2new}.} The
extension to non-vacuum models is straightforward and leads to the
same Hamiltonian for several standard matter models; the simplest way
to prove this is by following  the arguments in~\cite{Kijowski78}.

We show in~\cite{ChNagy2} that the vacuum Einstein equations together
with the decay conditions \eq{C4}-\eq{C5} and the asymptotic behavior
of the frame components of the $b$-Killing vector fields,
$$X^a=O(r)\;, \znabla_a X^b =O(r)\;,$$
guarantee that the integrals
\eq{toto} are convergent. 
 Further, it is easy to see that under the asymptotic conditions
\eq{C4}-\eq{C5} the volume integrals appearing in the variational
formulae in~\cite{ChAIHP} are convergent, the undesirable boundary
integrals in the variational formulae of~\cite{ChAIHP} vanish, so that
the integrals~\eq{toto} do indeed provide Hamiltonians on the space of
fields satisfying \eq{C4}-\eq{C5}. (Compare~\cite{HT} for an
alternative, essentially equivalent, Hamiltonian approach.)  This
singles out the charges~\eq{toto} amongst various alternative
expressions because Hamiltonians are uniquely defined, up to the
addition of a constant, on each path connected component of the phase
space.  The key advantage of the Hamiltonian approach is precisely
this uniqueness property, which does not seem to have a counterpart in
the Noether charge analysis~\cite{Trautman62} (\emph{cf.},
however~\cite{WaldZoupas,SilvaJulia:2000}), or in Hamilton-Jacobi type
arguments~\cite{YorkBrown}.

To define the integrals \eq{toto} we have fixed a model background
metric $b$, as well as an orthonormal frame as in \eq{frame}; this
last equation requires the corresponding coordinate system $(t,r,v^A)$
as in \eq{eq:a1}. Hence, the background structure required in our
analysis consists of a {\em background metric} and a {\em background
coordinate system}.  While this is
reasonably satisfactory from a Hamiltonian point of view, in which
each choice of background structure defines an associated phase space,
there is an essential
\emph{potential geometric ambiguity} in the integrals
\eq{toto} that arises as follows:  let $g$ be any metric such
that its frame components $g^{ab}$ tend to $\eta^{ab}$ as $r$ tends to
infinity, in such a way that the integrals $m(\hyp,g,b, X)$ given by
\eq{toto} (labelled by all the background Killing vector fields $X$ or
perhaps by a subset thereof) converge.  Consider another coordinate
system $(\hat t,\hat r,\hat v^A)$ with the associated background
metric $\hat b$:
\begin{eqnarray}\nonumber
  \hat b= -{a^{-2}(\hat r)} d\hat t^2 + a^2(\hat r) d\hat r^2 +\hat
r^2 \hat h\;,\qquad
  \hat h=h_{AB}(\hat v^C)d\hat v^A d\hat v  ^B\;,
  \label{A1hat}
\end{eqnarray}
 together with an associated frame $\hat e^a$,
\begin{eqnarray}
  \label{hframe}
\hat e_0 = a(\hat r) \partial_{\hat t}\;, \qquad \hat e_1 = {1 \over
  a(\hat r)  } \partial _{\hat r}\;,
\quad
  \hat e_{\A} = \frac {1 }{\hat r} \hat{\ce}_{\A}\;,
\end{eqnarray}
and suppose that in the new hatted coordinates the integrals defining
the charges $m(\hhyp,g,\hat b, \hat X)$ converge again. An obvious
way of obtaining such coordinate systems is to make a coordinate
transformation
\begin{equation}
  \label{eq:A5}
  t\to \hat t=t+\delta t\;,\ r\to \hat r=r+\delta r\;,\
v^A\to \hat v^A=v^A+\delta v^A\;,
\end{equation}
with $(\delta t, \delta r,\delta v^A)$ decaying sufficiently fast:
\begin{eqnarray} \nonumber
\hat{t} = t + O(r^{-1-\beta })\;, &
e_a(\hat{t}) = \ell\,\delta_a^0 + O(r^{-1-\beta })\;,
\\ \nonumber
\hat{r} = r + O(r^{1-\beta })\;, &
e_a(\hat{r}) = \frac{\delta_a^1}{\ell} + O(r^{1-\beta })\;,
\\ \label{coordtra} \hat{v}^A = v^A + O(r^{-1-\beta })\;, &
\; e_a(\hat{v}^A) = \delta_a^A + O(r^{-1-\beta }) \;,
\end{eqnarray} and with analogous  conditions on second
derivatives; this guarantees that the hatted analogue of
Equations~\eq{C4} and \eq{C5} will also hold. In~\cite{ChNagy2} we
prove that under coordinate transformations
\eq{coordtra} the integrals \eq{toto} remain unchanged:
$$m(\hyp,g,b,X)=m(\hhyp,g,\hat{b},\hat{X})\;.$$ Here, if
$X=X^\mu(t,r,v^A)\partial_\mu$, then the vector field $\hat X$ is
defined using the \emph{same} {functions} $X^\mu$ of the \emph{hatted}
variables.  The proof is a computation of the change of the integrand
of \eq{toto} under the change of coordinates \eq{eq:A5}; one notices
that all terms obtained by linearizing in $(\delta t,\delta r,\delta
v^A)$ can be written as a total divergence, and therefore give no
contribution; this is reminiscent of a divergence identity which is
used for a similar calculation for asymptotically flat space-times.
One finally checks that under the boundary conditions \eq{C4}-\eq{C5}
the correction terms give vanishing contribution in the limit
$r\to\infty$.

One does not expect all the requirements in \eq{C4}-\eq{C5} to be
necessary: for metrics which are asymptotically flat at $i_0$ it is
sufficient to impose conditions on the induced metric and the
extrinsic curvature of the hypersurface $\hyp$ to obtain a well
defined mass, and one expects the same to be the case here. However,
the decay rates imposed are sharp in the following sense:
 consider the metric $g=\hb$, with the hatted coordinates
defined as \begin{equation}
\label{ct}
\hat{r} = r+ \frac{\zeta}{r^{n/2-1}}
\;,\quad \hat{v}^A= v^{A}\;,
\end{equation}
where $\zeta$ is a constant.  Thus, if $\hb$ is the anti-de Sitter
metric, then $g$ is again anti-de Sitter metric in a different
coordinate system, which does not differ too much from the original
one; in particular the metric approaches asymptotically the standard
version of the anti-de Sitter metric in the unhatted
coordinates. Then the components of $g$ with respect to the ON
frame associated to the unhatted background $b$ satisfy
\eq{C4} with $\beta=\gamma=n/2$; a calculation shows that $g$ has
non-vanishing mass
integral \eq{toto} with respect to the unhatted background $b$.  The
same coordinate transformation  leads to a non-zero
Abbott-Deser~\cite{AbbottDeser} mass of $g=\hb$ with respect to $b$.

It should be stressed that we do not know \emph{a priori\/} that the
hatted coordinates are related to the unhatted ones by the simple
coordinate transformation \eq{eq:A5} with $(\delta t, \delta r,\delta
v^A)$ decaying as $r\to\infty$, or behaving in some controlled way ---
the behavior of $(\delta t, \delta r,\delta v^A)$ could in principle
be very wild.  The main technical result of~\cite{ChNagy2} is the proof
that this is not the case: under the hypothesis that $(M,h)$ is a
sphere with a round metric, or that $(M,h)$ has non-positive Ricci
tensor and constant Ricci scalar\footnote{Related results under
less restrictive conditions can be found in~\cite{ChNagy2}.} we show
that \emph{all coordinate transformations which leave $\hyp$ invariant
(so that $\hat t = t$) and that preserve the decay conditions
\eq{C4}-\eq{C5} are compositions of a map satisfying \eq{coordtra}
with an isometry of the background.} In order to obtain a geometric
invariant of $\hyp$ it remains thus to study the behavior of the
integrals \eq{toto} under isometries of $b$ preserving $\hyp$. If
$\Phi$ is such an isometry, the fact that $\ourU^{\alpha\beta}$ is a
tensor density immediately yields the formula \begin{equation}
\label{eq:l1} m(\hyp,\Phi^*g,b,(\Phi_*)^{-1}X) = m(\hyp,g,b,X)\;.
\end{equation}
(Here $\Phi^*$ is the pull-back, and $\Phi_*$ is the push-forward
map. When understood in a passive manner, $\Phi^*g$ is simply the
metric $g$ expressed in the new coordinates, while $(\Phi_*)^{-1}X$ is
the vector field $X$ expressed in the new coordinates.) This has the
effect that the \emph{integrals \eq{toto} are reshuffled amongst each
other under isometries of $b$, in a way determined by the
representation of the group of isometries of $b$ on the space of
Killing vectors,} and leads to the following:

  1. Let $M$ be the $n-1$ dimensional sphere $^{n-1}S$ with a
   round metric $h$,
normalized so that the
  substitution $k=1$ in \eq{eq:a1}  leads to a  metric $b$ which is the $n+1$
  dimensional anti-de Sitter metric.  The space $\cKS$ of $b$-Killing
  vector fields normal to $\hyp\cap\cU$ is spanned by vector fields
  $\Kbasis_{(\mu)}$ which on $\hyp$ take the form
  $\Kbasis_{(\mu)}=N_{(\mu)}e_0$, $\mu=0,\cdots,{n}$, where
  $N_{(0)}=\sqrt{{r^2\over\ell^2}+1}, 
  N_{(i)}={x^i\over\ell}$, 
and $x^i = r n^i$, $r$ being the
  coordinate which appears in \eq{eq:a1}, while
  $n^i\in{}^{n-1}S\subset\R^n$. The group $\mathit{Iso}(\hyp,b)$ of
  isometries $\Phi$ of $b$ which map $\hyp$ into $\hyp$ coincides with
  the homogeneous Lorentz group $O(1,n)$; it acts on $\cKS$ by
  push-forward. It can be shown that for each such $\Phi$ there exists
  a Lorentz transformation
  $\Lambda=(\Lambda^{(\alpha)}{}_{(\beta)}):\R^{n+1}\to\R^{n+1}$ so
  that for every $X=X^{(\alpha)}\Kbasis_{(\alpha)}\in\cKS$ we have
  $$(\Phi_* X)^{(\alpha)}=
  \Lambda^{(\alpha)}{}_{(\beta)}X^{(\beta)}\;.$$ We set
  $$m_{(\mu)}\equiv m\left(\hyp,g,b,{\Kbasis}_{(\mu)}\right)\;;$$ it
  follows that the number $$m^2(\hyp,g) = |\eta^{{(\mu)}{(\nu)}}
  m_{(\mu)} m_{(\nu)}|\;,$$ where
  $\eta^{{(\mu)}{(\nu)}}=\mathrm{diag}(-1,+1,\cdots,+1)$ is the
  Minkowski metric on $\R^{n+1}$, is an invariant of the action of
  $\mathit{Iso}(\hyp,b)$. Further, if we define $m(\hyp,g)$ to be
  positive if $m^{(\mu)}:=\eta^{{(\mu)}{(\nu)}} m_{(\nu)}$ is
  spacelike, and we take the sign of $m(\hyp,g)$ to coincide with
  that of $m^{(0)}$ if $m^{(\mu)}$ is timelike or null, then
  $m(\hyp,g)$ is invariant under the action of the group $\Isot$ of
  time-orientation preserving Lorentz transformations.  The number
  $m(\{t=0\},g)$ so defined coincides with the mass parameter $m$ of
  the Kottler (``Schwarzschild -- anti-de Sitter'') metrics in
  dimension $n+1=4$, and it is proportional to the parameter $m$ which
  occurs in the $(n+1)$-dimensional generalizations of the Kottler
  metrics~\cite{HorowitzMyers} \begin{eqnarray} g & = & -\left(1 -
  \frac {2m}{r^{n-2}} + {r^2\over \ell^2}\right) dt^2 + \left(1 -
  \frac {2m}{r^{n-2}} + {r^2\over \ell^2}\right)^{-1} dr^2
+r^2 h \;,\label{Kottler2}
\end{eqnarray}
with $h$ --- a round metric of scalar curvature $(n-1)(n-2)$ on a
$(n-1)$-dimensional sphere ({\em cf., e.g.},~\cite{CadeauWoolgar}).

Consider the remaining Killing vector fields
$L_{(\alpha)(\beta)}$ of $b$:
\begin{equation}
  \label{Lor2}
L_{(\alpha)(\beta)} = \eta_{(\alpha)(\sigma)}y^{(\sigma)} \frac{\partial~~}{\partial y^{(\beta)}}
-\eta_{(\beta)(\sigma)}y^{(\sigma)} \frac{\partial~~}{\partial y^{(\alpha)}}\;.
\end{equation} Here the coordinates $y^{(\alpha)}$ are the coordinates
on $\R^{n+1}$ obtained by isometrically embedding --- into $\R^{n+1}$
--- the
$\{t=0\}$ section of $n+1$-dimensional anti-de Sitter space-time
equipped with the Minkowski metric $\eta^{(\mu)(\nu)}$. It can be
checked that under the action of $\Iso$ the integrals
$$Q_{(\mu)(\nu)}\equiv m(\hyp,g,b,L_{(\mu)(\nu)})$$
transform as
the components of a two-covariant antisymmetric tensor. One then
obtains a geometric invariant of $\hyp$ by calculating
\begin{equation}
  \label{eq:ch}Q\equiv
  Q_{(\mu)(\nu)}Q_{(\alpha)(\beta)}\eta^{(\mu)(\alpha)}
\eta^{(\nu)(\beta)}\;.
\end{equation}
In dimension $3+1$ another
independent global geometric invariant is obtained from
\begin{equation}
  \label{eq:ch1}
  Q^*\equiv
Q_{(\mu)(\nu)}Q_{(\alpha)(\beta)}\epsilon^{(\mu)(\alpha)(\nu)(\beta)}\;.
\end{equation}
In higher dimensions further invariants are obtained by calculating
$\text{tr}(P^{2k})$, $2\le 2k\le (n+1)$, where
$P^{(\alpha)}{}_{(\beta)}= \eta^{(\alpha)(\mu)}Q_{(\mu)(\beta)}$.
(In this notation $Q$ given by \Eq{eq:ch} equals $\text{tr}(P^{2})$.)

2. Let $M$ be a compact $n-1$ dimensional manifold with a metric $h$
of constant scalar curvature and with non-positive Ricci tensor, and
let $b$ take the form \eq{eq:a1}, with $a(r)=1/\sqrt{r^2/\ell^2+k}$,
and with $k=0$ or $-1$ according to whether the Ricci scalar of $h$
vanishes or not.  We show in~\cite{ChNagy2} that for such metrics the
space of $b$-Killing vector fields normal to $\hyp$ consists of vector
fields of the form
$X(\lambda)= \lambda\partial_t, 
\lambda\in\R$,
and that
$$m(\hyp,g)\equiv m(\hyp,g,b,X(1))$$
is background independent, hence a geometric invariant\footnote{If the
scalar curvature of $h$ vanishes, then Einstein equations require the
constant $k$ in the metric to vanish. In that case there arises an
ambiguity in the definition of mass related to the possibility of
rescaling $t$ and $r$ without changing the form of the metric, which
rescales the mass. (This freedom does not occur when $k$ is non-zero.)
This ambiguity can be removed by arbitrarily choosing some
normalization for the $h$-volume of $\mn$, {\em e.g.} $4\pi$ in
dimension $n+1$=4.}.  Some other geometric invariants can be obtained
from the integrals \eq{toto} when Killing vectors which are not
necessarily normal to $\hyp$ exist, using invariants of the action of
the isometry group of $b$ on the space of Killing vectors.  If the
Ricci tensor of $\mn$ is strictly negative no other Killing vectors
exist. On the other hand, if $h$ is a flat torus, then each
$h$--Killing vector provides a global geometric invariant via the
integrals \eq{toto}.

The number $m(\hyp,g)$ defined in each case above is our proposal for
the geometric definition of total mass of $\hyp$ in $(\cM,g)$. We note
that its multiplicative normalization in $n+1=4$ dimensions is
determined by the requirement of the correct Newtonian limit when
$M=S^2$ and $\Lambda =0$, while the additive one is determined by
imposing that the background models have vanishing energy. In higher
dimensions it appears appropriate to keep the same multiplicative
factor for the Lagrangian $\mathcal L$, whenever a Kaluza-Klein reduction
applies.

The results described above can be reformulated in a purely
Riemannian context~\cite{ChHerzlich}.

It is natural to study the invariance of the mass when $\hyp$ is
allowed to move in $\cM$.  A complete answer would require
establishing an equivalent of our analysis of admissible coordinate
transformations in a space-time setting. The difficulties that arise
in the corresponding problem for asymptotically Minkowskian
metrics~\cite{Chmass} suggest that this might be a considerably more
delicate problem, which we plan to analyse in the future. It should be
stressed that this problem mixes two different issues, one being the
potential background dependence of \eq{toto}, another one being the
possibility of energy flowing in or out through the timelike conformal
boundary of space-time.

It should be pointed out that there exist several alternative
methods of defining mass in asymptotically anti-de Sitter
space-time --- using coordinate systems~\cite{BGH,GHHP83},
preferred foliations~\cite{GibbonsGPI}, generalized Komar
integrals~\cite{Magnon}, conformal
techniques~\cite{AshtekarDas,AshtekarMagnonAdS,Balasubramanian},
or \emph{ad-hoc} methods~\cite{AbbottDeser}; an extended
discussion can be found in~\cite[Section 5]{ChruscielSimon}. Each
of those approaches suffers from some \emph{potential}
ambiguities, so that the question of the geometric character of
the definition of mass given there arises as well.  Let us briefly
describe the relationship of the results presented here to some of
those works.  Consider, first, the Abbott-Deser
approach~\cite{AbbottDeser}, which seems to have been most often
used. A precise comparison is difficult to carry out because in
Ref.~\cite{AbbottDeser} the boundary conditions which should be
assumed are not spelled out in detail. As already pointed out, the
coordinate transformation~\eq{ct} gives a non-zero Abbott-Deser
mass to the anti-de Sitter metric, so the boundary conditions {\em
do} matter. Assuming that the authors of \cite{AbbottDeser} had in
mind boundary conditions which are at least as restrictive as our
conditions \eq{C4}-\eq{C5}, a straightforward but tedious
calculation shows that the Abbott-Deser mass coincides with the
Hamiltonian mass advocated here. One needs then to face the same
ambiguities as we do, and our results in~\cite{ChNagy2} can be
interpreted as proving the existence of a geometric invariant
which can be calculated using Abbott-Deser type integrals. It
should be stressed, however, that while the Hamiltonian approach
is universal and leads to unique --- up to a constant ---
functionals on each connected component of phase space, no results
about either uniqueness or universality of the approach of
\cite{AbbottDeser} are known to us.

Consider, next, the Hamiltonian approach of~\cite{HT}; here the
question of equivalence of the different Hamiltonian formalisms used
in~\cite{ChAIHP} and in~\cite{HT} arises. Those formalisms are
identical in spirit --- both are Hamiltonian --- but differ in various
details. We note that the boundary conditions
\eq{C4}-\eq{C5} are weaker than the conditions imposed
in~\cite{HT}.  It can be checked that under \eq{C4}-\eq{C5} the
formalism of~\cite{HT} is a special case of the geometric Hamiltonian
formalism of~\cite{ChAIHP}, when applied to asymptotically anti-de
Sitter space-times, so the Hamiltonian part of our analysis
in~\cite{ChNagy2} can be thought of as an extension of the results
of~\cite{HT} to larger phase spaces arising naturally in this
context.  An advantage of the geometric Hamiltonian formalism of
Kijowski and Tulczyjew is that it allows the use of both a manifestly
four-dimensional covariant formalism, and of a $3+1$ ADM one. The
{potential} ambiguities in a geometric definition of mass that occur
in~\cite{HT} are identical to the ones described above.

 As another example, we note the potential ambiguity in the mass
 defined by the conformal methods
 in~\cite{AshtekarDas,AshtekarMagnonAdS}, related to the possibility
 of existence of smooth conformal completions which are \emph{not}
 smoothly conformally equivalent. In those works one assumes existence
 of smooth completions, while our conditions would --- roughly
 speaking --- correspond to $C^{1,\alpha}$ conformal completions,
 $\alpha > 1/2$, hence our conditions
\eq{C4}-\eq{C5} are satisfied in the setup
of~\cite{AshtekarDas,AshtekarMagnonAdS} (compare~\cite{ChruscielSimon}
for a detailed discussion in the static case). It can be shown
(A.~Ashtekar, private communication) that the mass
of~\cite{AshtekarDas,AshtekarMagnonAdS} coincides with the
Abbott-Deser one; what has been said above implies that --- under the
asymptotic conditions of~\cite{AshtekarDas,AshtekarMagnonAdS} --- it
also coincides with the Hamiltonian mass described here. The results
proved in~\cite{ChNagy2} can be used to show that no inequivalent
conformal completions of the kind considered
in~\cite{AshtekarDas,AshtekarMagnonAdS} exist, establishing the
invariant character of the framework
of~\cite{AshtekarDas,AshtekarMagnonAdS}.

Let us, finally, turn our attention to the AdS/CFT motivated
definitions of mass ({\em cf.,
e.g.,\/}~\cite{AshtekarDas,Balasubramanian,EJM,MyersSTCE} and
references therein). It seems that there might be a belief in the
string community that, at least in odd space-time dimensions, there is
no systematic way to start from the Einstein-Hilbert action and arrive
unambiguously at conserved quantities. Our results in~\cite{ChNagy2}
show that such a belief is incorrect, in the space of metrics
asymptotic to the backgrounds considered here, as well as for those
which asymptote to the more general backgrounds considered
in~\cite{ChNagy2}. It should be pointed out, however, that the
Hamiltonian approach gives Hamiltonians which are only defined up to a
constant, which leaves ample room for non-zero ``Casimir energies'',
the occurrence of which might well be justified by other physical
considerations. On the other hand, in a Hamiltonian approach the only
natural choice of the additive constant seems to be zero. In any case,
it is not clear whether or not the AdS/CFT based approaches lead to
geometric invariants of hypersurfaces in space-times, in the sense
presented here.

We note that a similar problem for the ADM mass of asymptotically
flat initial data sets has been settled
in~\cite{Bartnik:mass,ChErice} (see also~\cite{Chmass}). Our
treatment in~\cite{ChNagy2} is a non-trivial extension to the
current setup of the methods of~\cite{ChErice}.

\textbf{Acknowledgements:} PTC wishes to thank
M.~Herzlich, S.~Ilias, A.~Polombo and A.El Soufi for useful
discussions. Suggestions from an anonymous referee for
improvements of the manuscript are acknowledged.

\bibliographystyle{amsplain}
\bibliography{
../../references/newbiblio,%
../../references/reffile,%
../../references/bibl,%
../../references/Energy,%
../../references/hip_bib,%
../../references/netbiblio}

\def\cprime{$'$}
\begin{thebibliography}{28}
\expandafter\ifx\csname natexlab\endcsname\relax\def\natexlab#1{#1}\fi
\expandafter\ifx\csname bibnamefont\endcsname\relax
  \def\bibnamefont#1{#1}\fi
\expandafter\ifx\csname bibfnamefont\endcsname\relax
  \def\bibfnamefont#1{#1}\fi
\expandafter\ifx\csname citenamefont\endcsname\relax
  \def\citenamefont#1{#1}\fi
\expandafter\ifx\csname url\endcsname\relax
  \def\url#1{\texttt{#1}}\fi
\expandafter\ifx\csname urlprefix\endcsname\relax\def\urlprefix{URL }\fi
\providecommand{\bibinfo}[2]{#2}
\providecommand{\eprint}[2][]{\url{#2}}

\bibitem[{\citenamefont{Chru\'sciel and Nagy}(2001)}]{ChNagy2}
\bibinfo{author}{\bibfnamefont{P.}~\bibnamefont{Chru\'sciel}} \bibnamefont{and}
  \bibinfo{author}{\bibfnamefont{G.}~\bibnamefont{Nagy}}
  (\bibinfo{year}{2001}), \bibinfo{note}{gr-qc/0110014}.

\bibitem[{\citenamefont{Chru\'sciel and Simon}(2001)}]{ChruscielSimon}
\bibinfo{author}{\bibfnamefont{P.}~\bibnamefont{Chru\'sciel}} \bibnamefont{and}
  \bibinfo{author}{\bibfnamefont{W.}~\bibnamefont{Simon}},
  \bibinfo{journal}{Jour.\ Math.\ Phys.} \textbf{\bibinfo{volume}{42}},
  \bibinfo{pages}{1779} (\bibinfo{year}{2001}), 
  \eprint{gr-qc/0004032}.

\bibitem[{\citenamefont{Friedrich}(1995)}]{Friedrich:adS}
\bibinfo{author}{\bibfnamefont{H.}~\bibnamefont{Friedrich}},
  \bibinfo{journal}{Jour.\ Geom.\ and Phys.} \textbf{\bibinfo{volume}{17}},
  \bibinfo{pages}{125} (\bibinfo{year}{1995}).

\bibitem[{\citenamefont{K{\'a}nn{\'a}r}(1996)}]{Kannar:adS}
\bibinfo{author}{\bibfnamefont{J.}~\bibnamefont{K{\'a}nn{\'a}r}},
  \bibinfo{journal}{Class.\ Quantum Grav.} \textbf{\bibinfo{volume}{13}},
  \bibinfo{pages}{3075} (\bibinfo{year}{1996}).

\bibitem[{\citenamefont{Chru\'sciel}(1985)}]{ChAIHP}
\bibinfo{author}{\bibfnamefont{P.}~\bibnamefont{Chru\'sciel}},
  \bibinfo{journal}{Ann. Inst. H. Poincar\'e} \textbf{\bibinfo{volume}{42}},
  \bibinfo{pages}{267} (\bibinfo{year}{1985}).

\bibitem[{\citenamefont{Kijowski and Tulczyjew}(1979)}]{KijowskiTulczyjew}
\bibinfo{author}{\bibfnamefont{J.}~\bibnamefont{Kijowski}} \bibnamefont{and}
  \bibinfo{author}{\bibfnamefont{W.}~\bibnamefont{Tulczyjew}},
  \emph{\bibinfo{title}{A Symplectic Framework for Field Theories}}, vol.
  \bibinfo{volume}{107} of \emph{\bibinfo{series}{Lecture Notes in Physics}}
  (\bibinfo{publisher}{Springer}, \bibinfo{address}{New York, Heidelberg,
  Berlin}, \bibinfo{year}{1979}).

\bibitem[{\citenamefont{Kijowski}(1978)}]{Kijowski78}
\bibinfo{author}{\bibfnamefont{J.}~\bibnamefont{Kijowski}},
  \bibinfo{journal}{Gen. Rel. Grav.} \textbf{\bibinfo{volume}{9}},
  \bibinfo{pages}{857} (\bibinfo{year}{1978}).

\bibitem[{\citenamefont{Henneaux and Teitelboim}(1985)}]{HT}
\bibinfo{author}{\bibfnamefont{M.}~\bibnamefont{Henneaux}} \bibnamefont{and}
  \bibinfo{author}{\bibfnamefont{C.}~\bibnamefont{Teitelboim}},
  \bibinfo{journal}{Commun.\ Math.\ Phys.} \textbf{\bibinfo{volume}{98}},
  \bibinfo{pages}{391} (\bibinfo{year}{1985}).

\bibitem[{\citenamefont{Trautman}(1962)}]{Trautman62}
\bibinfo{author}{\bibfnamefont{A.}~\bibnamefont{Trautman}}, in
  \emph{\bibinfo{booktitle}{Gravitation: an introduction to current research}},
  edited by \bibinfo{editor}{\bibfnamefont{L.}~\bibnamefont{Witten}}
  (\bibinfo{publisher}{Wiley}, \bibinfo{year}{1962}).

\bibitem[{\citenamefont{Wald and Zoupas}(2000)}]{WaldZoupas}
\bibinfo{author}{\bibfnamefont{R.}~\bibnamefont{Wald}} \bibnamefont{and}
  \bibinfo{author}{\bibfnamefont{A.}~\bibnamefont{Zoupas}},
  \bibinfo{journal}{Phys. Rev.} \textbf{\bibinfo{volume}{D61}},
  \bibinfo{pages}{084027 (16 pp.)} (\bibinfo{year}{2000}),
  \bibinfo{note}{gr-qc/9911095}.

\bibitem[{\citenamefont{Julia and Silva}(2000)}]{SilvaJulia:2000}
\bibinfo{author}{\bibfnamefont{B.}~\bibnamefont{Julia}} \bibnamefont{and}
  \bibinfo{author}{\bibfnamefont{S.}~\bibnamefont{Silva}},
  \bibinfo{journal}{Class. Quantum Grav.} \textbf{\bibinfo{volume}{17}},
  \bibinfo{pages}{4733} (\bibinfo{year}{2000}), \eprint{gr-qc/0005127}.

\bibitem[{\citenamefont{Brown and {York, Jr.}}(1993)}]{YorkBrown}
\bibinfo{author}{\bibfnamefont{J.}~\bibnamefont{Brown}} \bibnamefont{and}
  \bibinfo{author}{\bibfnamefont{J.}~\bibnamefont{{York, Jr.}}},
  \bibinfo{journal}{Phys. Rev.} \textbf{\bibinfo{volume}{D47}},
  \bibinfo{pages}{1407} (\bibinfo{year}{1993}).

\bibitem[{\citenamefont{Abbott and Deser}(1982)}]{AbbottDeser}
\bibinfo{author}{\bibfnamefont{L.}~\bibnamefont{Abbott}} \bibnamefont{and}
  \bibinfo{author}{\bibfnamefont{S.}~\bibnamefont{Deser}},
  \bibinfo{journal}{Nucl.\ Phys.} \textbf{\bibinfo{volume}{B195}},
  \bibinfo{pages}{76} (\bibinfo{year}{1982}).

\bibitem[{\citenamefont{Horowitz and Myers}(1999)}]{HorowitzMyers}
\bibinfo{author}{\bibfnamefont{G.}~\bibnamefont{Horowitz}} \bibnamefont{and}
  \bibinfo{author}{\bibfnamefont{R.}~\bibnamefont{Myers}},
  \bibinfo{journal}{Phys. Rev.} \textbf{\bibinfo{volume}{D59}},
  \bibinfo{pages}{026005 (12 pp.)} (\bibinfo{year}{1999}),
  \eprint{hep-th/9808079}.

\bibitem[{\citenamefont{Cadeau and Woolgar}(2001)}]{CadeauWoolgar}
\bibinfo{author}{\bibfnamefont{C.}~\bibnamefont{Cadeau}} \bibnamefont{and}
  \bibinfo{author}{\bibfnamefont{E.}~\bibnamefont{Woolgar}},
  \bibinfo{journal}{Class.\ Quantum Grav.} pp. \bibinfo{pages}{527--542}
  (\bibinfo{year}{2001}), \bibinfo{note}{gr-qc/0011029}.

\bibitem[{\citenamefont{Chru\'sciel and Herzlich}()}]{ChHerzlich}
\bibinfo{author}{\bibfnamefont{P.}~\bibnamefont{Chru\'sciel}} \bibnamefont{and}
  \bibinfo{author}{\bibfnamefont{M.}~\bibnamefont{Herzlich}},
  \bibinfo{note}{math.DG/0110035}.

\bibitem[{\citenamefont{Chru\'sciel}(1988)}]{Chmass}
\bibinfo{author}{\bibfnamefont{P.}~\bibnamefont{Chru\'sciel}},
  \bibinfo{journal}{Commun.\ Math.\ Phys.} \textbf{\bibinfo{volume}{120}},
  \bibinfo{pages}{233} (\bibinfo{year}{1988}).

\bibitem[{\citenamefont{Boucher et~al.}(1984)\citenamefont{Boucher, Gibbons,
  and Horowitz}}]{BGH}
\bibinfo{author}{\bibfnamefont{W.}~\bibnamefont{Boucher}},
  \bibinfo{author}{\bibfnamefont{G.}~\bibnamefont{Gibbons}}, \bibnamefont{and}
  \bibinfo{author}{\bibfnamefont{G.}~\bibnamefont{Horowitz}},
  \bibinfo{journal}{Phys.\ Rev.\ D} \textbf{\bibinfo{volume}{30}},
  \bibinfo{pages}{2447} (\bibinfo{year}{1984}).

\bibitem[{\citenamefont{Gibbons et~al.}(1983)\citenamefont{Gibbons, Hawking,
  Horowitz, and Perry}}]{GHHP83}
\bibinfo{author}{\bibfnamefont{G.}~\bibnamefont{Gibbons}},
  \bibinfo{author}{\bibfnamefont{S.}~\bibnamefont{Hawking}},
  \bibinfo{author}{\bibfnamefont{G.}~\bibnamefont{Horowitz}}, \bibnamefont{and}
  \bibinfo{author}{\bibfnamefont{M.}~\bibnamefont{Perry}},
  \bibinfo{journal}{Commun.\ Math.\ Phys.} \textbf{\bibinfo{volume}{88}},
  \bibinfo{pages}{295} (\bibinfo{year}{1983}).

\bibitem[{\citenamefont{Gibbons}(1999)}]{GibbonsGPI}
\bibinfo{author}{\bibfnamefont{G.}~\bibnamefont{Gibbons}},
  \bibinfo{journal}{Class.\ Quantum Grav.} \textbf{\bibinfo{volume}{16}},
  \bibinfo{pages}{1677} (\bibinfo{year}{1999}).

\bibitem[{\citenamefont{Magnon}(1985)}]{Magnon}
\bibinfo{author}{\bibfnamefont{A.}~\bibnamefont{Magnon}},
  \bibinfo{journal}{Jour.\ Math.\ Phys.} \textbf{\bibinfo{volume}{26}},
  \bibinfo{pages}{3112} (\bibinfo{year}{1985}).

\bibitem[{\citenamefont{Ashtekar and Das}(2000)}]{AshtekarDas}
\bibinfo{author}{\bibfnamefont{A.}~\bibnamefont{Ashtekar}} \bibnamefont{and}
  \bibinfo{author}{\bibfnamefont{S.}~\bibnamefont{Das}},
  \bibinfo{journal}{Class.\ Quantum Grav.} \textbf{\bibinfo{volume}{17}},
  \bibinfo{pages}{L17} (\bibinfo{year}{2000}), \bibinfo{note}{hep-th/9911230}.

\bibitem[{\citenamefont{Ashtekar and Magnon}(1984)}]{AshtekarMagnonAdS}
\bibinfo{author}{\bibfnamefont{A.}~\bibnamefont{Ashtekar}} \bibnamefont{and}
  \bibinfo{author}{\bibfnamefont{A.}~\bibnamefont{Magnon}},
  \bibinfo{journal}{Class.\ Quantum Grav.} \textbf{\bibinfo{volume}{1}},
  \bibinfo{pages}{L39} (\bibinfo{year}{1984}).

\bibitem[{\citenamefont{Balasubramanian and Kraus}(1999)}]{Balasubramanian}
\bibinfo{author}{\bibfnamefont{V.}~\bibnamefont{Balasubramanian}}
  \bibnamefont{and} \bibinfo{author}{\bibfnamefont{P.}~\bibnamefont{Kraus}},
  \bibinfo{journal}{Commun.\ Math.\ Phys.} \textbf{\bibinfo{volume}{208}},
  \bibinfo{pages}{413} (\bibinfo{year}{1999}), \eprint{hep-th/9902121}.

\bibitem[{\citenamefont{Emparan et~al.}(1999)\citenamefont{Emparan, Johnson,
  and Myers}}]{EJM}
\bibinfo{author}{\bibfnamefont{R.}~\bibnamefont{Emparan}},
  \bibinfo{author}{\bibfnamefont{C.}~\bibnamefont{Johnson}}, \bibnamefont{and}
  \bibinfo{author}{\bibfnamefont{R.}~\bibnamefont{Myers}},
  \bibinfo{journal}{Phys. Rev.} \textbf{\bibinfo{volume}{D60}},
  \bibinfo{pages}{104001 (14 p.)} (\bibinfo{year}{1999}),
  \bibinfo{note}{hep-th/9903238}.

\bibitem[{\citenamefont{Myers}(1999)}]{MyersSTCE}
\bibinfo{author}{\bibfnamefont{R.}~\bibnamefont{Myers}},
  \bibinfo{journal}{Phys. Rev.} \textbf{\bibinfo{volume}{D60}},
  \bibinfo{pages}{046002 (12 p.)} (\bibinfo{year}{1999}),
  \bibinfo{note}{hep-th/9903203}.

\bibitem[{\citenamefont{Bartnik}(1986)}]{Bartnik:mass}
\bibinfo{author}{\bibfnamefont{R.}~\bibnamefont{Bartnik}},
  \bibinfo{journal}{Comm. Pure Appl. Math.} \textbf{\bibinfo{volume}{39}},
  \bibinfo{pages}{661} (\bibinfo{year}{1986}).

\bibitem[{\citenamefont{Chru\'sciel}(1986)}]{ChErice}
\bibinfo{author}{\bibfnamefont{P.}~\bibnamefont{Chru\'sciel}}, in
  \emph{\bibinfo{booktitle}{Topological Properties and Global Structure of
  Space--Time}}, edited by
  \bibinfo{editor}{\bibfnamefont{P.}~\bibnamefont{Bergmann}} \bibnamefont{and}
  \bibinfo{editor}{\bibfnamefont{V.}~\bibnamefont{de~Sabbata}}
  (\bibinfo{publisher}{Plenum Press}, \bibinfo{address}{New York},
  \bibinfo{year}{1986}), pp. \bibinfo{pages}{49--59}, \bibinfo{note}{{URL}
  \url{http://www.phys.univ-tours.fr/~piotr/scans}}.

\end{thebibliography}

\end{document}